\documentclass[letterpaper]{article}
\usepackage{uai2020}
\usepackage[margin=1in]{geometry}

\usepackage{subfig}

\usepackage{fancyhdr}       

\usepackage[natbibapa]{apacite}
\usepackage{amsthm}
\usepackage{amsmath}
\usepackage{amssymb}
\usepackage[toc]{appendix}
\usepackage{mathtools}
\newtheoremstyle{personal}
{5pt}
{5pt}
{}
{8pt}
{\scshape}
{:}
{.5em}
{}

\newtheorem{definition}{Definition}[section]

\newtheorem{theorem}[definition]{Theorem}

\newtheorem{assumption}{Assumption}

\DeclareMathOperator*{\Pa}{Pa}

\DeclareMathOperator*{\Des}{De}

\DeclareMathOperator*{\biota}{\boldsymbol{\iota}}
\DeclareMathOperator*{\biotabar}{\bar{\boldsymbol{\iota}}}

\DeclareMathOperator*{\etabar}{\bar{\boldsymbol{\eta}}}

\DeclareMathOperator*{\bn}{\textbf{n}}
\DeclareMathOperator*{\bX}{\textbf{X}}
\DeclareMathOperator*{\bTheta}{\boldsymbol{\Theta}}

\usepackage{tikz}
\usetikzlibrary{arrows,snakes,backgrounds,decorations.pathreplacing}
\tikzstyle{place}=[circle,minimum width=15pt,draw=blue!50,fill=blue!20,thick]
\tikzstyle{transition}=[rectangle,minimum width=15pt,draw=black!50,fill=black!20,thick]
\tikzstyle{node}=[scale=.85,minimum width=15pt,draw=white,fill=white,thick]
\tikzstyle{snode}=[scale=.85,circle,minimum width=32pt,draw=black,fill=white,thick]
\tikzstyle{white}=[scale=.85,circle,minimum width=25pt,draw=white,fill=white,thick]

\usepackage{times}

\title{Evaluation of Causal Structure Learning Algorithms via Risk Estimation}

\author{} 

\author{ {\bf Marco F. Eigenmann} \\
Seminar f\"ur Statistik\\
ETH Zurich\\
Zurich, Switzerland\\
\And
{\bf Sach Mukherjee}  \\
German Center for Neurodegenerative\\
Diseases (DZNE)\\
Bonn, Germany \\
\And
{\bf Marloes H. Maathuis}   \\
Seminar f\"ur Statistik\\
ETH Zurich\\
Zurich, Switzerland\\
}

\begin{document}

\maketitle

\begin{abstract}
Recent years have seen many advances in methods for causal structure learning from data. The empirical assessment of such methods, however, is much less developed. Motivated by this gap, we pose the following question: how can one assess, in a given problem setting, the practical efficacy of one or more causal structure learning methods? We formalize the problem in a decision-theoretic framework, via a notion of expected loss or risk for the causal setting. We introduce a theoretical notion of causal risk as well as sample quantities that can be computed from data, and study the relationship between the two, both theoretically and through an extensive simulation study. 
Our results provide an assumptions-light framework for assessing causal structure learning methods that can be applied in a range of practical use-cases.
\end{abstract}

\section{INTRODUCTION}

\label{Subsection: Introduction}

Causal structure learning has seen many recent developments and the literature is growing rapidly. A range of algorithms have been developed under different assumptions. These include, among others, PC \citep{Spirtes2000}, FCI \citep{Spirtes2000}, GES \citep{ChickeringGES}, LiNGAM \citep{Shimizu2006}, MMHC \citep{Tsamardinos2006}, GIES \citep{Hauser2012}, RFCI \citep{Colombo2012}, FCI+ \citep{Claassen2013}, order-independent PC \citep{ColomboMaathuis2014}, rank PC \citep{Harris2013}, CAM \citep{Buehlmann14}, ICP \citep{Peters2015b}, AGES \citep{Eigenmann2017}, ARGES \citep{Nandy2015}, LGES \citep{Frot2017}, and MRCL \citep{Hill2019}.

The majority of these papers contain theoretical guarantees for the developed algorithms as well as simulation studies showing their empirical performance, including comparisons with competing algorithms. In simulation studies estimated graphs can be compared to the ground truth, e.g.
using the Structural Hamming Distance (SHD) \citep{Acid2011, Tsamardinos2006} or the more causal oriented Structural Intervention Distance (SID) \citep{Peters2015}. Alternatively, one can consider particular features related to a graph like the total causal effect between two nodes \citep{Maathuis2009}.



However, by design and scope, simulation studies have some key limitations. In particular, good performance in a simulation does not imply good performance on a given real-world problem, since a real data-generating system may violate model assumptions in such a way as to strongly affect the relevant output. While model assumptions may be tested in principle using various statistical tools, causal assumptions in particular can be difficult if not impossible to test directly.
Thus, in practice, given a data set obtained from a specific system, it remains challenging to choose among algorithms, or to assess a given algorithm. Some work has been done to fill this theoretical-empirical gap \citep{Sachs2005, Mooij2013, Hill2016}. This usually involves very interesting and challenging interdisciplinary collaborations which allow to infer a ground truth to which causal methods can be compared.

In this paper, we address the question of evaluating causal structure learning  in a more general sense. Our approach is rooted in a decision-theoretic view of causal structure learning and leads to procedures that could be applied generally, wherever suitable data is available. 
Thus, our goal is not to propose a new approach to estimate causal graphs, but a new approach to assess existing methods in a problem-specific manner. 

The remainder of the paper is organized as follows. We begin with a problem statement, clarifying precisely the question we seek to address. We then propose a notion of {\it causal risk} as well as corresponding sample quantities that could be used to assess causal risk in practice, and study their relationship. We then show results from a large simulation study, covering more than 40,000 data-generating regimes, aimed at investigating the practical performance of the criteria we propose.

\section{PROBLEM STATEMENT AND SUMMARY OF CONTRIBUTIONS} \label{Subsection: The problem setting}

{\it Problem statement.} We aim to 
evaluate the performance of causal structure learning algorithms on a given data set containing some observational and some interventional data. Ideally, we would wish to be able to select the best performing algorithm (among those considered) for the specific problem setting. This problem statement acknowledges that different methods may perform better or worse in specific problem settings (this will becomes precise via the decision-theoretic framework we introduce below).
We want to construct an assumptions-light framework and therefore will only assume that the data come from a structural equation model (SEM; see Definition~\ref{Definition: SEM}), without imposing many restrictions on the SEM. In particular, we will not assume joint independence or a particular distribution for the noise terms, nor will we assume acyclicity.

{\it Why assessment of causal learning is hard.}  
It is useful to consider at a high-level why empirical assessment of causal structure learning methods is nontrivial and different from familiar non-causal tasks in machine learning and statistics. 
In typical non-causal tasks (such as classification/regression or probabilistic modelling, e.g. via non-causal graphical models) performance measures rooted in classical sampling theory make sense, because the core assumption is that all data -- current and future -- share the same probability model. In contrast, a causal model encodes a {\it collection of distributions} (e.g. arising from different interventions on the system; see Def.\ ~\ref{Definition: SEM}~and~\ref{Definition: Interventions} below) and this limits the scope of familiar sampling theory-based approaches to assessment. 

It is instructive to consider this difference with an example. In a regression problem, assuming that a fixed and unique distribution underlies the data permits (i) the use of residuals
as proxies for the statistical noise (that can be used to check assumptions about the noise, via e.g., Tukey-Anscombe or QQ-plots)
and (ii) the use of various cross-validation-type methods to test prediction accuracy. 
For variable selection, candidate procedures can be evaluated using likelihood methods applied to selected variables, and similar strategies can be used for non-causal model selection in general. In contrast, for causal problems, the fact that one is dealing with a collection of potentially very different distributions does not allow the use of sampling techniques like cross-validation in a straightforward way. Moreover, in causal systems a good model needs to go beyond out-of-sample performance and cope with (potentially strongly) out-of-distribution scenarios from which no data may be available.


{\it Summary of contributions.}
Our main contributions are as follows. (i)
We show how causal structural learning can be viewed through a decision-theoretic lens, and propose a notion of causal loss that allows assessment via expected loss or risk. (ii) We study the question of estimating causal risk from data and 
propose assumptions-light risk estimation procedures that can be used in practice using interventional data. (iii) We study the behaviour of our procedures in theory and in an extensive simulation study spanning more than 40,000 data-generating regimes.

%
%
The core idea of our approaches is to exploit information given by the interventional data to generalize the performance of the algorithms to other unseen interventions. 
To this end, we use simple statistical tests that do not require the same types of assumptions as causal structure learning methods, and whose output allows us to estimate a useful notion of causal risk. 
Causal relationships 
have been estimated in a risk minimization framework \citep{Arjovsky2019}, 
and held-out interventional data has been used in applications \citep{Hill2016},
but to the best of our knowledge, the present work is the first formal risk estimation framework for causal structure learning.


\section{CAUSAL RISK}\label{Section: Causal Risk}

\subsection{PRELIMINARIES AND NOTATION}\label{Subsection: Preliminaries}

We associate vertices in a directed graph $G$ with variables and say that variable $X_j$ is a parent of $X_i$ if the directed edge $X_j \rightarrow X_i$ is included in $G$. We say that $X_j$ is a descendant of $X_i$ in $G$ if there is a directed path $X_i \rightarrow \ldots \rightarrow X_j$ in $G$. We denote the set of parents and descendants of $X_i$ in $G$ by $\Pa(G,i)$ and $\Des(G,i)$, respectively. We let $[p]:= \{1,\dots,p\}$.

We assume that the data come from a structural equation model (SEM).
\begin{definition}\label{Definition: SEM}
A structural equation model is a system of equations $\mathcal{S}=\{S_1,\ldots ,S_p\}$ on a set of variables $\{X_1,\ldots, X_p\}$:
\begin{equation}\label{Equation: Def SEM}
S_i: X_i \leftarrow f_{i}(X_{\Pa(G,i)},\varepsilon_i), \qquad i\in [p],
\end{equation}
where $G$ denotes the directed graph associated with the SEM, and the noise terms $\varepsilon_1,
\dots,\varepsilon_p$ have mean $0$, and finite variance.
\end{definition}
The assignment arrow in Equation \eqref{Equation: Def SEM} emphasizes the causal relationship between its left and  right hand sides. In other words, $\mathcal S_i$ is understood as the generating mechanism of $X_i$. A SEM can be represented by a directed graph $G$, where for any pair $(X_i,X_j)$, there is a directed edge $X_j \rightarrow X_i$ if $X_j$ is involved in structural equation $\mathcal S_i$, that is, 
if $j \in \Pa(G,i)$. 
Thus, a direct edge represents a direct effect and $X_{\Pa(G,i)}$ are the direct causes of $X_i$.

Some causal modelling frameworks require acyclicity of the  graph $G$ and independence of the noise terms, but we do not assume this here.

\begin{definition}\label{Definition: Interventions}
An intervention on a set of nodes $\{X_i: i\in I\}$ is modelled by replacing the respective structural equations by 
$$\widetilde{S}_i: X_i \leftarrow  \widetilde{f}_{i}(X_{\Pa(\widetilde G,i)},\widetilde{\varepsilon}_i), \qquad \quad i\in I,$$
where $\widetilde f_i$, $\widetilde{G}$ and $\widetilde \varepsilon_i$ are respectively the functional form, the directed graph, and the noise variable under the intervention. The structural equations for $i {\notin} I$ remain unchanged. 
\end{definition}

We assume that we have $n_0$ i.i.d.\ observations from an unknown SEM (see Def.~\ref{Definition: SEM}), as well as some i.i.d.\ observations from different interventions (see Def.~\ref{Definition: Interventions}). 
For ease of exposition, we assume that we only have single interventions, meaning that an intervention affects exactly one node or structural equation. 
We denote by $\biota \subseteq [p]$ the collection of nodes on which we have data from single interventions, and we let $n_i$, $i\in \biota$ be the corresponding sample sizes.  
It will turn out to be convenient to use an augmented set
$\biotabar$ that includes the observational data, denoted by a $0$, i.e. $\biotabar = \{0\} \cup \biota$. The corresponding sample sizes are denoted by $\bn=(n_i: i\in \biotabar)$. The total sample size is $N=\sum_{i\in \biotabar} n_i$.

We denote by $\bTheta$ all parameters necessary to fully represent the SEM and its interventions. For instance, the necessary parameters for a linear Gaussian SEM would be the edge weights, the means and variances of the noise terms in the original regime, and the new weights, means, and variances under the interventions in $\biota$.
To emphasize that the true underlying graph representing our SEM is unknown, we will from now on denote it by $G^\star$. Further, we denote by $G^\star_{\bTheta}$ the multivariate distributions that arise with the parameters in $\bTheta$. 

Data coming from the SEM and its interventions are denoted by $\bX_{\bn,\biotabar,\bTheta} \sim G^\star_{\bTheta}$, where $\bX_{\bn,\biotabar,\bTheta} \in \mathbb{R}^{N \times p}$. We emphasize that this represents a sample from a {\it collection} of $|\biota|+1$ multivariate distributions arising from the underlying causal system.
To simplify notation, we will in the sequel suppress the dependence on $\bn$ and $\bTheta$, and indicate only the set of interventions $\biotabar$. That is, we write $\bX_{\biotabar}$ instead of $\bX_{\bn,\biotabar,\bTheta}$. We also consider leaving out data on certain interventions, considering only a subset $\etabar \subset \biotabar$. In that case, we write $\bX_{\etabar}$. We consider either all or none of the samples under a certain intervention, so that the sample sizes corresponding to $\etabar$ equal the corresponding entries of $\bn$.

\subsection{THE ORACLE RISK FUNCTION}\label{Subsection: Oracle risk function}

We first define a theoretical notion of risk that involves the true graph $G^*$. We emphasize that 
this theoretical quantity cannot be computed in practice. We will consider practically applicable estimates of the theoretical risk in  Section \ref{Subsection: Risk estimators}. 

Let $\hat{H}$ be a causal structure learning algorithm that returns a graph. Let $\hat H(\bX_{\biotabar})$ denote the graph that is returned when the algorithm is applied to data set $\bX_{\biotabar}$. The general form of the risk function we propose is the following:
\begin{align}\label{Equation: General risk function}
R_{\biotabar}(\hat{H}) &= \mathbb{E}_{\bX_{\biotabar}\sim G^\star_{\bTheta}}\left[ L(G^\star, \textstyle\hat{H}(\bX_{\biotabar})) \right],
\end{align}
where $L$ is a loss function acting on a pair of directed graphs.

Note that this notion of risk is problem-specific in the sense that it quantifies the finite sample efficacy of method $\hat{H}$ in the specific context defined by the system $G^\star_{\bTheta}$. This allows for the possibility that a given method may do well in some settings but not in others.

There are several choices to be made in order to define a concrete risk function to study. In particular, we must define a loss function. As we will motivate in Section~\ref{Subsubsection: Remarks risk function} below, we will consider a node-wise loss function that compares the {\it descendants} of nodes of the graphs $G^\star$ and $\hat{H}(\bX_{\biotabar})$. 
In particular, we consider the following special case of Equation~\eqref{Equation: General risk function}
\begin{equation}\label{Equation: Concrete risk function}
\begin{multlined}
R_{\biotabar}^J(\hat{H}) = \\
 \mathbb{E}_{\bX_{\biotabar} \sim G^{\star}_{\bTheta} } \frac{1}{p} \sum_{i=1}^p \left( J(\Des(G^\star,i), \Des(\hat{H} \textstyle(\bX_{\biotabar}),i) \right), 
\end{multlined}
\end{equation}
where $J(A,B)$ denotes the Jaccard distance between two sets $A$ and $B$, defined as $J(A,B) = 1 - \frac{\vert A \cap B\vert}{\vert A \cup B\vert}$ if $A\cup B \neq \emptyset$ and 0 otherwise. 

We note that due to identifiability issues, some causal structure learning algorithms do not return a directed graph, but for example a partially directed graph. In that case, $\Des(\hat{H} \textstyle(\bX_{\biotabar}),i)$ should be adapted, depending on the interpretation of the output graph. For example, if the output is a completed partially directed acyclic graph (CPDAG; \citealp{Andersson1997, Chickering2002}), the descendants of $X_i$ can be replaced by the set of possible descendants of $X_i$, i.e., the set of nodes $X_j$ for which there is a partially directed path from $X_i$ to $X_j$. We will make this concrete in Section \ref{Section: Simulations}. To set things up, however, we will first omit these issues and think in terms of directed graphs.

\subsubsection{Remarks on the Oracle Risk Function}\label{Subsubsection: Remarks risk function}
A key challenge in working with Equation~\eqref{Equation: General risk function} is the presence of $G^\star$. In particular, if $\biota \subsetneq \left[ p \right]$ there is at least one intervention that is of interest but under which we have no data. This limits the utility of standard likelihood-based and cross-validation-type approaches since we cannot sample from the unobserved intervention. Therefore, we need an approach that does not rely on explicit knowledge of $G^\star$.

When looking at general interventions as defined in Definition~\eqref{Definition: Interventions} we can quickly see that the observational and interventional data distinguish themselves on the descendants of the intervened node. Indeed, the distribution of all descendants of $X_i$ is potentially different. It is therefore quite natural to use the descendants as a feature that captures information that is causally relevant, estimable and that may generalize to unobserved interventions.


\subsection{RISK ESTIMATORS}\label{Subsection: Risk estimators}

\subsubsection{Descendant Estimation}\label{Subsection: Descendants estimation}

Since the true underlying graph $G^*$ is unknown, we must construct an estimate for $\Des(G^\star,i)$ in Equation \eqref{Equation: Concrete risk function}. This is virtually impossible for $i {\notin} \biota$. For $i {\in} \biota$, however, it is feasible by comparing the observational data, $\bX_{\{0\}} \! \in \! \mathbb{R}^{n_0 \times p}$, to the data under the intervention on node $i$, $\bX_{\{i\}} \! \in \! \mathbb{R}^{n_i \times p}$. In particular, we can compare the $j$th column of $\bX_{\{0\}}$ to the $j$th column of $\bX_{\{i\}}$. If these are significantly different, we conclude that the intervention on $X_i$ has affected $X_j$, and hence that $X_j$ is a descendant of $X_i$. This approach only works if the aggregated effect of directed paths from $X_i$ to $X_j$ in $G^*$ does not cancel, if there is at least one such path. The latter is related to the common faithfulness assumption \citep{Spirtes2000}.

Concretely, for each intervention node $i\in \biota$, we conduct $p-1$ two-sample tests, comparing the $j$th column of $\bX_{\{0\}} \in \mathbb{R}^{n_0 \times p}$ to the $j$th column of $\bX_{\{i\}} \in \mathbb{R}^{n_i \times p}$, for $j\in [p]\setminus\{i\}$. If a significant difference is found, $X_j$ is declared to be an estimated descendant of $X_i$. With a slight abuse of notation we denote the set of estimated descendants of node $X_i$ by $\widehat{\Des}(\bX_{\biotabar},i)$. This will substitute $\Des(G^\star,i)$ in the risk estimators.

We note that our framework is general, in the sense that we do not specify the type of interventions (e.g., do-interventions or shift interventions) nor the statistical tests that should be used. We will require, however, that one has some knowledge of the type of interventions, so that one can conduct appropriate statistical tests to detect the descendants. We refer to Section~\ref{Subsection: Descendants estimation Simulation study} for the concrete example we used in the simulation study.

\subsubsection{The Main Risk Estimator}\label{Subsubsection: The main risk estimator}

We consider three risk estimators: the first two (Equation~\eqref{Equation: Naive risk estimator}~and~\eqref{Equation: CV risk estimator}) serve as auxiliary risk estimators and are needed to construct the third one (Equation~\eqref{Equation: Main risk estimator}), which is the main risk estimator we propose.

Ideally, we would like to compute 
\begin{equation}\label{Equation: Naive oracle risk estimator}
\begin{multlined}
\hat{R}^J_{\biotabar,\text{oracle}}\textstyle (\hat{H},\bX_{\biotabar})=\\
\displaystyle \frac{1}{p}\sum_{i=1}^p \textstyle J(\Des(G^\star,i),\Des(\hat{H}(\bX_{\biotabar}),i)),
\end{multlined}
\end{equation}
where $\Des(G^\star,i)$ should be replaced by some estimate of it. As discussed in Section~\ref{Subsection: Descendants estimation}, however, we can only estimate $\Des(G^\star,i)$ by $\widehat{\Des}(\bX_{\biotabar},i)$ for $i \in \biota$.\footnote{The ``hat" on ${R}^J_{\biotabar,\text{oracle}}$ in equation \eqref{Equation: Naive oracle risk estimator} is used to indicate that this is a random variable that depends on the data $\bX_{\biotabar}$.} Restricting the node-wise sum to these terms and scaling appropriately, we obtain our first risk estimator: 
\begin{equation}\label{Equation: Naive risk estimator}
\begin{multlined}
\hat{R}^J_{\biotabar,\text{naive}}\textstyle (\hat{H},\bX_{\biotabar})=\\
\displaystyle \frac{1}{\vert \biota \vert}\sum_{i\in \biota} \textstyle J(\widehat{\Des}(\bX_{\biotabar},i),\Des(\hat{H}(\bX_{\biotabar}),i)).
\end{multlined}
\end{equation}
Equation~\eqref{Equation: Naive risk estimator} is a natural first step since it contains everything we can estimate. Indeed, for $i \notin \biota$, $\widehat{\Des}(\bX_{\biotabar},i)$ is not defined and cannot be defined in a reasonable and natural way. Hence, in order to sum over all nodes as done in Equation~\eqref{Equation: Naive oracle risk estimator}, we would have to make outside assumptions that cannot be supported by data. This should be avoided and Section~3 of the supplementary material illustrates failure cases that can arise when doing this.

We next define a risk estimator that focuses more explicitly on the out-of-distribution aspect.
This risk estimator uses a leave-one-out cross-validation-like approach on the intervention nodes and is defined as
\begin{equation}\label{Equation: CV risk estimator}
\begin{multlined}
\hat{R}^J_{\biotabar,CV}\textstyle (\hat{H},\bX_{\biotabar}) = \\
\displaystyle \frac{1}{\vert \biota \vert} \sum_{i\in \biota}\textstyle J(\widehat{\Des}(\bX_{\{0,i\}},i),\Des(\hat{H}(\bX_{\biotabar\setminus\{i\}}),i)).
\end{multlined}
\end{equation}
This expression uses a \textit{distributional splitting} scheme in which the interventional data is split into two disjoint groups.\footnote{Note that $\widehat{\Des}(\bX_{\biotabar},i) = \widehat{\Des}(\bX_{\{0,i\}},i)$.}  This is fundamentally different from randomly splitting the sample as in classical cross-validation.

In words, Equation \eqref{Equation: CV risk estimator} does the following. It applies the algorithm under investigation, $\hat{H}$, $\vert\biota\vert$ times. For each $i\in \biota$, we pass to the algorithm all observational data and the interventional data corresponding to interventions in $\biota\setminus\{i\}$, $\bX_{\biotabar\setminus\{i\}}$, and determine the descendants of $X_i$ in the resulting graph, $\Des(\hat{H}(\bX_{\biotabar\setminus\{i\}}),i)$. At the same time, we use the observational data and interventional data corresponding to the intervention on $i$, $\bX_{\{0,i\}}$, to estimate the descendants of $X_i$ using some two-sample tests, yielding $\widehat{\Des}(\bX_{\{0,i\}},i)$. By comparing these two estimated sets of descendants, we emulate the evaluation of the performance of $\hat{H}$ on unseen interventions. Finally, this is averaged over $i\in \biota$. 

Our main risk estimator $\hat{R}^J_{\biotabar,w}(\hat{H},\bX_{\biotabar})$ combines Equation~\eqref{Equation: Naive risk estimator}~and~\eqref{Equation: CV risk estimator} as a weighted sum, where the weights, $\frac{\vert \biota \vert}{p}$ and $\frac{p-\vert \biota \vert}{p}$, correspond to the proportion of nodes with and without interventions, respectively:
\begin{equation}\label{Equation: Main risk estimator}
\begin{multlined}
\textstyle \hat{R}^J_{\biotabar,w}(\hat{H},\bX_{\biotabar}) =\\
\textstyle \frac{\vert \biota \vert}{p} \hat{R}_{\biotabar,\text{naive}}^J(\hat{H},\bX_{\biotabar}) + \frac{p-\vert \biota \vert}{p}\textstyle \hat{R}_{\biotabar,CV}^J(\hat{H},\bX_{\biotabar}).
\end{multlined}
\end{equation}
This estimator balances both aspects, the in-distribution performance on seen interventions through $\hat{R}_{\biotabar,\text{naive}}^J(\hat{H},\bX_{\biotabar})$, and the out-of-distribution performance on unseen interventions through $\textstyle \hat{R}_{\biotabar,CV}^J(\hat{H},\bX_{\biotabar})$.

\subsection{PROPERTIES OF THE MAIN RISK ESTIMATOR}\label{Subsection: Properties of the risk estimators}


We now investigate under which circumstances we can, in expectation, rank two algorithms correctly.
Concretely, for a given setting characterized by $G^*_{\bTheta}$ and $\biota$ and two causal structure learning methods $\hat{H}_1$ and $\hat{H}_2$,
we investigate when 
the difference $R_{\biotabar}^J(\hat{H}_1) - R_{\biotabar}^J(\hat{H}_2)$ in the oracle risks (defined in Equation~\eqref{Equation: Concrete risk function}) and the corresponding expected difference with respect to our proposed risk estimator
$\hat{R}^J_{\biotabar,w}$, i.e.,
$$\mathbb{E}_{\bX_{\biotabar} \sim G^\star_{\bTheta}}\left[ \hat{R}^J_{\biotabar,w}\textstyle (\hat{H}_1,\bX_{\biotabar}) - \hat{R}^J_{\biotabar,w}\textstyle (\hat{H}_2,\bX_{\biotabar}) \right],$$
have the same sign. Of course, this task should be easier if the difference in oracle risks is larger. Our results will therefore depend on 
$$ \delta := \left|R_{\biotabar}^J(\hat{H}_1) - R_{\biotabar}^J(\hat{H}_2)\right|.$$

Since our focus is on the performance of the risk estimator, we will assume for simplicity that the descendant estimation has oracle performance. 


\begin{assumption}(Oracle performance of the descendant estimation)\label{Assumption: Consistent test statistics}

We assume that the descendant estimation via $\widehat{\Des}(\bX_{\biotabar},i)$ achieves oracle performance with respect to $G^\star_{\bTheta}$ and $\biota$: $\widehat{\Des}(\bX_{\biotabar},i) = \Des(G^\star,i)$ for all $\bX_{\bar\biota} \sim G^*_{\bTheta}$ and $i \in \biota$.
\end{assumption}

Assumption \ref{Assumption: Consistent test statistics} is essentially one of correctness of the statistical decisions in a classical testing sense. It allows us to write $\Des(G^\star,i)$ instead of $\widehat{\Des}(\bX_{\biotabar},i)$ for $i\in \biota$ in the risk estimators. 

Next, we need to link the expected estimated difference in performance based on the cross-validation risk estimator 
\begin{align}\label{Equation: Exp Diff CV Risk}
\mathbb{E}_{\bX_{\biotabar} \sim G^\star_{\bTheta}}\left[
\hat{R}^J_{\biotabar,CV}\textstyle (\hat{H}_1,\bX_{\biotabar})  - \hat{R}^J_{\biotabar,CV}\textstyle (\hat{H}_2,\bX_{\biotabar}) \right],
\end{align} 
using the seen interventions in $\biota$, to the true difference in performance of $\hat H_1$ and $\hat H_2$ on unseen interventions on nodes $i\notin \biota$,
\begin{align}\label{Equation: Exp Or Diff Unseen Int}
& \mathbb{E}_{\bX_{\biotabar} \sim G^\star_{\bTheta}} \Bigg[ \frac{1}{p-\vert \biota \vert}\sum_{i \notin \biota} \left( \textstyle J(\Des(G^\star,i),  \textstyle \Des(\hat{H}_1(\bX_{\biotabar}),i))\right. \notag \\ 
& \qquad \qquad \quad \left. - \textstyle J(\Des(G^\star,i),  \textstyle \Des(\hat{H}_2(\bX_{\biotabar}),i))\right) \Bigg].
\end{align}
This link must only be made if there actually are unseen interventions, i.e., $|\biota|<p$. 

Now consider two algorithms $\hat{H}_1$ and $\hat{H}_2$ and a setting defined by $G^\star_{\bTheta}$ and $\biota$ with $|\biota| < p$. We say that $\hat H_1$ and $\hat H_2$ satisfy expected relative $\delta$-performance on unseen interventions with respect to $G^\star_{\bTheta}$ and $\biota$ if 
${\textstyle \big| \eqref{Equation: Exp Diff CV Risk} - 
          \eqref{Equation: Exp Or Diff Unseen Int} \big| < \frac{p}{p-|\biota|}\delta}.$
This will serve as our second assumption. 

\begin{assumption}(Expected relative $\delta$-performance on new interventions)\label{Definition: Expected relative delta performance}

We assume that algorithms $\hat{H}_1$ and $\hat{H}_2$ satisfy expected relative $\delta$-performance on new interventions with respect to $G^\star_{\bTheta}$ and $\biota$ (with $|\biota| < p)$.
\end{assumption}

The expected relative $\delta$-performance on unseen interventions incorporates the following two components:
(i) The performance of the algorithms using all data $\bX_{\bar\biota}$ must be similar to the performance of the algorithms when the data on one intervention is omitted, i.e., using $\bX_{\bar\biota \setminus \{i\}}$ for $i\in \biota$, as is done in the cross-validation risk estimator. 
(ii) The performance of the algorithms on the seen interventions in $\biota$ must be representative of the algorithms' performance on unseen interventions. 

The latter point is most important. We note that it is not testable; it is in essence an extrapolation type assumption that allows us to generalize the performance of the cross-validation risk estimator from seen to unseen interventions, and it becomes increasingly strong as the number of interventions decreases. 

We now obtain the following theorem. Its proof can be found in Section~1 of the supplementary material.

\begin{theorem}\label{Lemma: Same sign main risk estimator}
Consider a setting defined by $G^*_{\bTheta}$ and $\biota$ ($|\biota|>1$) and two algorithms $\hat H_1$ and $\hat H_2$ with oracle risk difference $\delta = \big\vert R_{\biotabar}^J(\hat{H}_1) - R_{\biotabar}^J(\hat{H}_2) \big\vert$. 

If $|\biota|=p$ and $\hat H_1$ and $\hat H_2$ satisfy Assumption \ref{Assumption: Consistent test statistics} with respect to $G^\star_{\bTheta}$ and $\biota$, or if $|\biota|<p$ and and $\hat H_1$ and $\hat H_2$ satisfy Assumptions \ref{Assumption: Consistent test statistics} and \ref{Definition: Expected relative delta performance} with respect to $G^\star_{\bTheta}$ and $\biota$, then 
$$R_{\biotabar}^J(\hat{H}_1) - R_{\biotabar}^J(\hat{H}_2)$$
and
$$\mathbb{E}_{\bX_{\biotabar} \sim G^\star_{\bTheta}}\left[ \hat{R}^J_{\biotabar,w}\textstyle (\hat{H}_1,\bX_{\biotabar}) - \hat{R}^J_{\biotabar,w}\textstyle (\hat{H}_2,\bX_{\biotabar}) \right]$$
have the same sign.
\end{theorem}

%
%
%


This result says that, under the given assumptions, the expected estimated difference in risk of the two algorithms has the correct sign. This is reassuring, as this is a property that a sensible risk estimator should have. Theorem \ref{Lemma: Same sign main risk estimator} does not guarantee, however, that the algorithms are correctly ranked for a particular realization of the data. In the simulations in Section \ref{Section: Simulations} we will assess the practical performance of our proposed risk estimator. 

We note that the assumption of expected relative $\delta$-performance acts as expected in the following ways: (i) If the true difference in oracle risks $\delta$ is larger, then the condition is weaker. This makes sense, since we have more ``room for error" in the estimation before we flip the sign. (ii) If we have interventional data on a large proportion of the nodes, then the factor $\frac{p}{p-|\biota|}$ is large and the condition also becomes weaker. This can be explained by the fact that in this case the weighted risk estimator gives a large weight to the naive risk estimator and only a small weight to the cross-validation risk estimator. In other words, there is less out-of-distribution assessment to be done.

\section{SIMULATION STUDY}\label{Section: Simulations}

In this Section, we empirically investigate the behaviour of the proposed risk estimation procedure via a simulation study. The basic strategy is as follows: we simulate data from many different known SEMs, that is, from many distributions $G^\star_{\bTheta}$.
In each such regime, since we know the true graph $G^\star$, 
we can compute the oracle risk (defined in Section~\ref{Subsection: Oracle risk function}) and thereby empirically assess agreement with our proposed risk estimator. The goal is to investigate behaviour in a range of finite-sample settings, where all estimation is done using available data, as would be the case in practical applications.

In line with the theoretical framework, we want to understand whether it is possible to distinguish, in an entirely data-driven manner, whether a certain method is more effective than another. This question is most urgent when two methods differ greatly in oracle risk, since then an incorrect choice means high cost or regret. 
Hence, we require a set of approaches for learning structure 
that would be collectively expected to span a range of performance levels. To this end we included both principled causal methods and simple non-causal estimators (that were expected to perform poorly). 
We note that we are not surveying all potentially useful methods for any particular setting, and acknowledge that many valid algorithms for the simulated settings have not been considered.
We emphasize that the goal of the simulation is not to offer guidance on specific methods that might work well in specific settings, but to study risk estimation {\it per se}.

\subsection{CONSIDERED SETTINGS}\label{Subsection: Considered settings}
In order to cover a wide variety of settings, we
 sampled a large parameter space, in a similar manner to the simulation study in \cite{Heinze2018}. The settings are defined below, and all parameters were sampled uniformly from the given ranges. 
We considered settings in which the statistical tests (for descendant estimation) were appropriate, as well as settings which violated assumptions of the tests.


\emph{The causal graph.} The causal graph was taken to be a directed acyclic graph, obtained by choosing a causal order on an Erd\H{o}s-R\'{e}nyi graph with $p\in \{25,50,100,200\}$ nodes and expected neighborhood size $\text{ENS} \in \{1.5, 2.5\}$.

\emph{The SEM.} We took SEMs of the following form
\begin{align*}
&\mathcal{S}_{i} : X_i \leftarrow \sum_{j=1}^{i-1} f(b_{ji},X_j) + \varepsilon_i, \; i \in [p],
\end{align*}
where the variables are assumed to be in a causal order, and $b_{ji} \neq 0$ if and only if $X_j \rightarrow X_i$ in $G^\star$. Here the nonzero $b_{ji}$'s are sampled uniformly from $[-3,-1] \cup [1,3]$. For a given SEM, the link functions are all of the same type, and are chosen to be either linear, 
or sigmoidal (expressions appear below).

The noise variables are taken to be jointly independent. For a given SEM, they all have the same type of distribution, which is chosen to be either Gaussian or lognormal, both with mean zero. The noise variance is set to 1 for source nodes. For non-source nodes, the noise variance and the edge weights were scaled to obtain variables with unit variance and a signal to noise ratio of 5. For details we refer to Section~2 of the supplementary material.

\emph{Interventions.}
We consider two types of interventions: Shift and Do-and-Shift. Both have a mean-shift component which is set to $5$ throughout, meaning that the noise distribution of an intervened node is shifted by 5. For Do-and-Shift, we additionally delete all incoming edges and set the noise variance of the node to $1$. 

For a given SEM, the interventions were either all Shift interventions, or all Do-and-Shift interventions, and each node had a probability $P_{\biota}$ to be intervened upon, independently of each others, where $P_{\biota}\in\{0.1, 0.2,5 0.5, 1\}$. 

\emph{Data.} For each SEM, one data set was generated, consisting of both observational and interventional data. The interventional sample sizes $n_{i}$, $i {\in} \biota$ for a SEM were taken to be identical and equal to $n_{int} \in \{10,100,1000\}$. The observational sample size $n_{0}$ was set to $\max(n_{int}, 100)$.

The parameter space is summarized below. Additional details regarding the simulations can be found in Section~2 of the supplementary material.
\begin{enumerate}
	\item Causal graph
	\begin{itemize}
		\item Number of variables $p\in\{25,50,100,200\}$
		\item Expected neighborhood size $\text{ENS}\in \{1.5,2.5\}$
	\end{itemize}		
	
	\item SEM
	\begin{itemize}
		
		\item Non-zero edge weights $b_{ji} \in [ -3,-1] \cup [ 1,3]$
		
		\item Link functions $f(b_{ji}, X_j)$ in the SEM:
		\begin{itemize}
			\item linear: $b_{ji} X_j$
			\item sigmoidal: $b_{ji}(\frac{10}{(1+\exp(-0.65*X_j))}-5)$
		\end{itemize}
		
		\item Noise distribution:
		\begin{itemize}
		\item $\mathcal{N}(0,1)$
			\item log-normal$(0,1)-e^{0.5}$
		\end{itemize}
		\item The noise and edge weights were scaled to obtain variables with unit variance and a signal to noise ratio of 5 for non-source nodes.  

	\end{itemize}
	
	\item Interventions
	\begin{itemize}
	\item Probability for each node to be intervened upon (independently): $P_{\biota}\in\{0.1, 0.2, 0.5, 1\}$
	\item Intervention type: Shift or Do-and-Shift, both with a mean shift of $5$
	\end{itemize}	
	
    \item Data
	\begin{itemize}
	
		\item Sample sizes of the interventions: $n_{i} = n_{int}$ for all $i\in \biota$, with $n_{int} \in\{10,100,1000\}$, and observational data size $n_{0} = \max(n_{int}, 100)$
		
	\end{itemize}

\end{enumerate}

For each sampled setting we generated a data set and ran GES and GIES, giving estimates $\widehat{GES}$ and $\widehat{GIES}$, respectively. These are principled causal algorithms that were run with standard settings as implemented in the \texttt{R}-package \texttt{pcalg} \citep{Kalisch2012}. GIES is expected to perform better because it is geared towards settings with interventional data. We also considered graphs based on the Pearson correlation (expected to perform worse than the causal methods). In particular, we considered the graph with the same expected neighborhood size as $G^\star$, using in essence an oracle cut-off to the matrix of (absolute) correlations and also, for comparison, 
the almost empty graph consisting of only one undirected edge between the nodes with the largest correlation coefficient. We denote these two algorithms by $\widehat{ACor}$ and $\widehat{Empty}$, respectively\footnote{We note that $\widehat{ACor}$ uses oracle information (true ENS of $G^\star$); this is intended to provide a simple point of comparison with correct sparsity, but with performance expected to be below an appropriate causal algorithm but better than random guessing.}.

For all algorithms we interpreted undirected edges as possibly directed edges. Accordingly, we replaced descendant sets in the risk estimator by their corresponding possible descendant sets. 
In order to have more stable and meaningful results we imposed the following two conditions on the settings used.
First, we considered only data sets that contain at least two interventions, so that the cross-validation based estimator can use some interventional data. 
Second, the intervened nodes were required to have at least three descendants in total with respect to $G^\star$. Note that this is a condition on the number of descendants (not out-degree) and serves to limit the occurrence of situations with zero true positives, in which case the Jaccard loss can only take the values $0$ or $1$. 

\subsection{DESCENDANT ESTIMATION}\label{Subsection: Descendants estimation Simulation study}

We used two-sample t-tests for a difference in mean between observational and interventional data to obtain $\widehat{\Des}(\bX_{\biota},i) = \widehat{\Des}(\bX_{\{0,i\}},i)$, which is required in the building blocks \eqref{Equation: Naive risk estimator} and \eqref{Equation: CV risk estimator} of our main risk estimator \eqref{Equation: Main risk estimator}. For each intervention we test for a difference in mean between the observational and interventional data of every node but the intervened one. 
The cut-off was computed with a multiplicity correction based on an empty graph, and under the assumption of Gaussian noise terms. Please see Section~2.1 of the supplementary material for details. 

We expect the t-tests to be a good choice in the linear Gaussian SEM. For the linear log-normal and sigmoidal case, behavior should still be reasonable for large sample sizes due to the central limit theorem. In the sigmoidal log-normal case, however, we can run into some issues. Indeed, a non-linear transformation applied to a non-symmetric distribution introduces an artificial mean, in the sense that one can obtain a mean shift that is not due to an intervention. This likely yields more false positives in the descendant estimation. This latter scenario is meant to assess the sensitivity of our framework to incorrect descendant estimation. (But we note that in a real-world use-case, one could center every node based on the observational data and avoid this problem.)


\subsection{RESULTS}\label{Section: Results}


Thus, we investigate agreement in ranking under true and estimated risk in a range of data-generating regimes. 
Figure~\ref{Figure: Summary tol01 main risk GESvsACor} shows a summary of results (additional results shown in supplementary material). Here a difference in true risk refers to the quantity 
${\hat{R}^J_{\biotabar,\text{oracle}}\textstyle (\hat{H}_1,\bX_{\biotabar}) - \hat{R}^J_{\biotabar,\text{oracle}}\textstyle (\hat{H}_2,\bX_{\biotabar})}$,
where $\hat{H}_1, \hat{H}_2$ are the two causal structure learning methods being compared. A difference in estimated risk refers to the 
corresponding quantity obtained from the risk estimator:
${\hat{R}^J_{\biotabar,\text{w}}\textstyle (\hat{H}_1,\bX_{\biotabar}) - \hat{R}^J_{\biotabar,\text{w}}\textstyle (\hat{H}_2,\bX_{\biotabar})}$.
We emphasize that risk estimation uses only the finite sample data generated in the specific example.

{\it How to read the plot.}
The plot should be read as follows. Each cell in the upper panel corresponds to a specific data-generating regime (defined by combinations of the factors listed above). The specific regime is indicated by the labels shown and the color indicates the difference in true risk. 
Corresponding cells in the lower panel refer to the same regimes and 
show
how often estimated differences in risks agreed in sign with the true risk difference (for the respective regime). 
For example, the very top left cell is the regime with a linear SEM with $p$=200 nodes, Normal noise, Do-and-Shift interventions where the probability for each node to be intervened upon was 0.1, and sample size 10. The colorbar shows that, in this case, the difference in true risk is large. The corresponding cell in the lower panel shows that the estimated difference in risk agrees in sign with the true risk difference. 

\emph{The choice of the algorithms shown.}
We chose to show the results on the two algorithms 
$\widehat{GES}$ and $\widehat{ACor}$,
because their comparison shows patterns that 
appear to hold more generally (see additional plots in Section~3.1 of the supplementary material).
We note that some other pairs, such as for example $\widehat{GES}$ and $\widehat{GIES}$
(both principled causal estimators)
showed even better results.


\begin{figure*}[!ht]
\centering
\includegraphics[width = \textwidth, height = 0.41\textwidth]{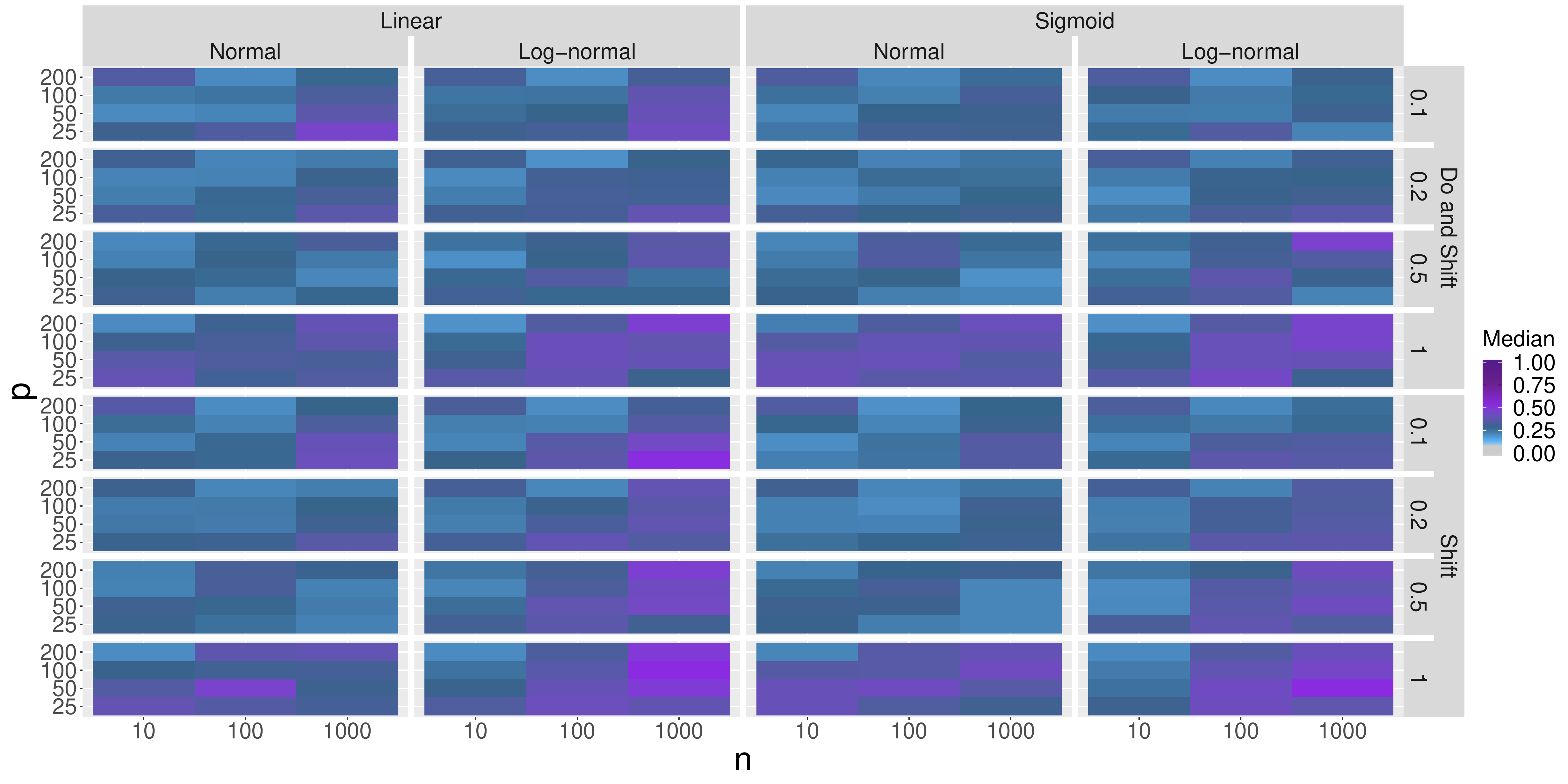}

\includegraphics[width = \textwidth, height = 0.41\textwidth]{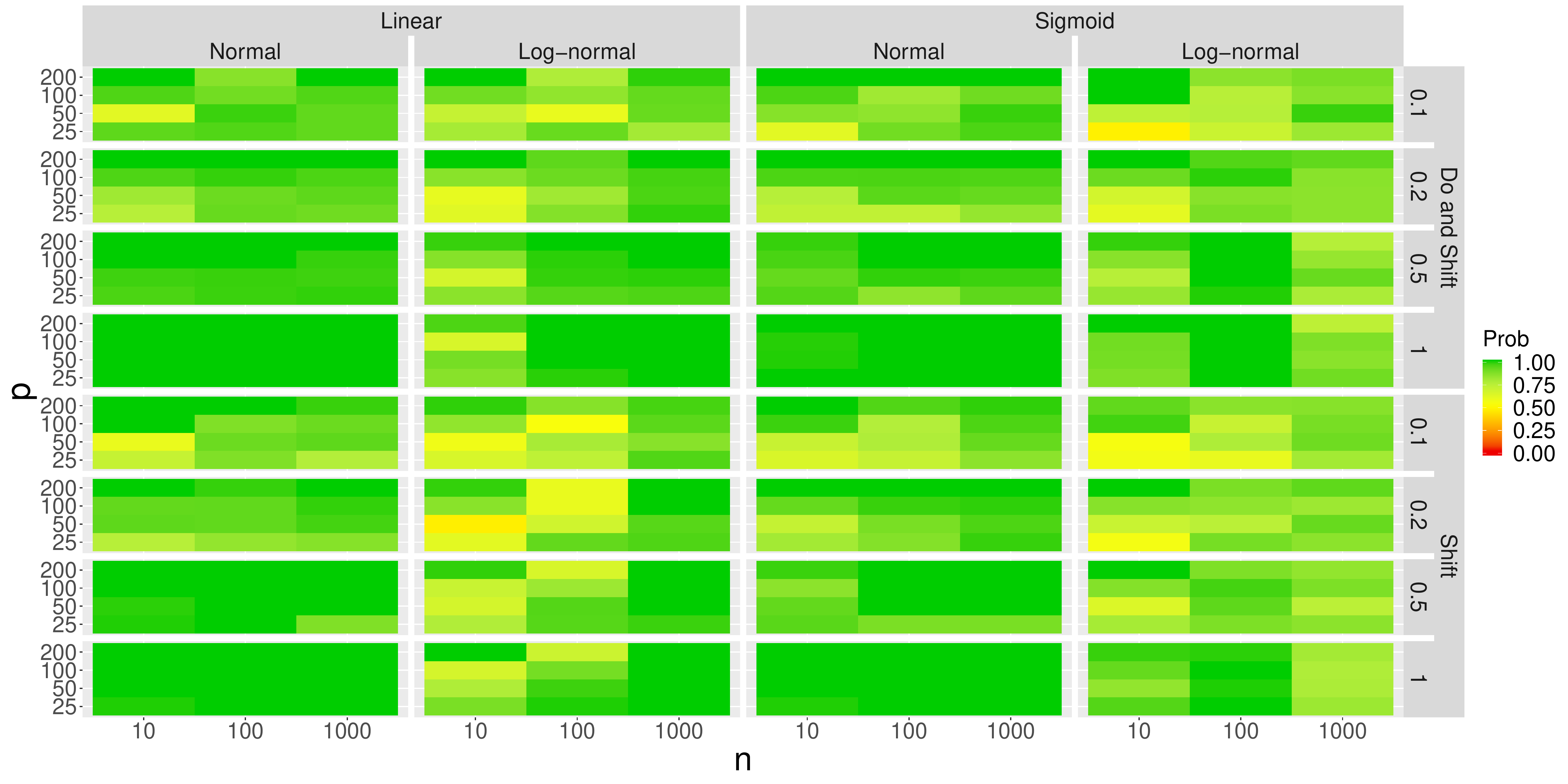}

\caption{For both the top and bottom panel: Each cell corresponds to a simulation setting, characterized by $p$ (left vertical axis), $n$ (bottom horizontal axis), the link functions and error distribution of them SEM (top horizontal axis) and the type and probability of an intervention (right vertical axis). Upper panel: median difference of true risk 
between methods {\it GES} and {\it ACor} (see text)
for different settings. A small value (blue) represents a more difficult situation to evaluate (since the true risks are then similar).
Lower panel: empirical probabilities for how often the corresponding difference in estimated risk 
agrees in sign with the difference of true risk for different settings. A large value (green) means that the risk estimator performed well in this sense. 
In each cell we consider only settings for which the true risks differ by at least $0.1$. A cell is left gray if less than $3$ settings are available.
}
\label{Figure: Summary tol01 main risk GESvsACor}
\end{figure*}






{\it Key insights.}
We can see that the our main risk estimator shows good performance, i.e. agreement with the 
ranking under the oracle, 
under many different settings and across a range of true performance differences (from light blue to violet in the upper panel of the Figure). In the more favourable regimes it is often the case that all signs are estimated correctly. These regimes also include examples with a lower percentages of interventions (as shown). Since all risk estimation was done using only the finite sample, regime-specific data (as would be available in a real-world application), these results suggest that our approach, or modifications of it, could be used to assess causal structure learning in practice. Sample size and descendant estimation plays an important role (see also Section~3.3 of the supplementary material). 
The size of the graph positively impacts risk estimator performance. As expected, more interventions help, as this makes the assumptions of Theorem~\ref{Lemma: Same sign main risk estimator} weaker.
In the sigmoidal log-normal case we see poor performance in the small graph and small sample setting. This is likely due to model mis-specification with respect to the testing approach. However, we note that this is an issue within the scope of classical statistical testing and could be resolved in practice by appropriate testing methods.

We emphasize that the results in Figure~\ref{Figure: Summary tol01 main risk GESvsACor} concern risk estimation in specific data-generating regimes. In our framework, the performance of a method is assessed in the context of a data-generating regime. This means that selecting the ``best" algorithm should not be interpreted as a general superiority statement. Further, good performance is defined with respect to the specific risk function used here. Our framework could be adapted to different definitions of descendants and different choices of loss functions more tailored to particular research interests (that might lead to different rankings).

Finally, we note that for small differences it can be more instructive to estimate the actual performance in terms of Equation~\eqref{Equation: Concrete risk function}, that is, whether both methods have a low or high causal risk.
In Section~3.2 of the supplementary material we show some results concerning this task.

\section{CONCLUSIONS}\label{Section: Conclusion}

We proposed a formal risk estimation framework for the evaluation of causal structure learning algorithms. This new framework represents a practical tool to facilitate the use, assessment and interpretation of causal structure learning methods.


We showed theoretically and empirically that the proposed approach -- that involves only quantities that can be computed from available data -- 
is indeed able to agree with a ranking of methods that would be possible given access to a true underlying causal graph. 
In a simulation study, covering a wide range of data-generating regimes, we found that often a large majority of signs are estimated correctly. Importantly, these scenarios are not limited to settings with many interventions, large oracle differences, or particular algorithms. This suggests that our approach, or extensions of it, have the potential to allow truly practical assessment of causal structure learning. As the field of causal structure learning continues to advance, we think questions around problem-setting-specific empirical assessment will become ever more important in real-world applications. 

Further work will be needed to weaken the assumptions regarding the statistical tests and to make this framework as general as possible. Moreover, different uses of this framework, for instance to tune the parameters of causal algorithms like PC and GES, can expand its scope and lead to additional interesting results and applications.

\subsubsection*{Acknowledgements}

MFE was supported by the Swiss SNF grant 200021\_172603. SM is a member of the German Bundesministerium f\"ur Bildung und Forschung (BMBF) consortium ``MechML".

\subsubsection*{References}
\renewcommand{\refname}{}
\bibliographystyle{apacite}
\bibliography{Bib}

\begin{thebibliography}{}

\bibitem [\protect \citeauthoryear {%
Acid%
\ \BBA {} de Campos%
}{%
Acid%
\ \BBA {} de Campos%
}{%
{\protect \APACyear {2003}}%
}]{%
Acid2011}
\APACinsertmetastar {%
Acid2011}%
\begin{APACrefauthors}%
Acid, S.%
\BCBT {}\ \BBA {} de Campos, L\BPBI M.%
\end{APACrefauthors}%
\unskip\
\newblock
\APACrefYearMonthDay{2003}{}{}.
\newblock
{\BBOQ}\APACrefatitle {Searching for {B}ayesian Network Structures in the Space
  of Restricted Acyclic Partially Directed Graphs} {Searching for {B}ayesian
  network structures in the space of restricted acyclic partially directed
  graphs}.{\BBCQ}
\newblock
\APACjournalVolNumPages{J.\ Artif.\ Intell.\ Res.}{18}{}{445-490}.
\PrintBackRefs{\CurrentBib}

\bibitem [\protect \citeauthoryear {%
Andersson%
, Madigan%
\BCBL {}\ \BBA {} Perlman%
}{%
Andersson%
\ \protect \BOthers {.}}{%
{\protect \APACyear {1997}}%
}]{%
Andersson1997}
\APACinsertmetastar {%
Andersson1997}%
\begin{APACrefauthors}%
Andersson, S\BPBI A.%
, Madigan, D.%
\BCBL {}\ \BBA {} Perlman, M\BPBI D.%
\end{APACrefauthors}%
\unskip\
\newblock
\APACrefYearMonthDay{1997}{}{}.
\newblock
{\BBOQ}\APACrefatitle {A characterization of {M}arkov equivalence classes for
  acyclic digraphs} {A characterization of {M}arkov equivalence classes for
  acyclic digraphs}.{\BBCQ}
\newblock
\APACjournalVolNumPages{Ann.\ Stat.}{25}{}{505--541}.
\PrintBackRefs{\CurrentBib}

\bibitem [\protect \citeauthoryear {%
Arjovsky%
, Bottou%
, Gulrajani%
\BCBL {}\ \BBA {} Lopez-Paz%
}{%
Arjovsky%
\ \protect \BOthers {.}}{%
{\protect \APACyear {2019}}%
}]{%
Arjovsky2019}
\APACinsertmetastar {%
Arjovsky2019}%
\begin{APACrefauthors}%
Arjovsky, M.%
, Bottou, L.%
, Gulrajani, I.%
\BCBL {}\ \BBA {} Lopez-Paz, D.%
\end{APACrefauthors}%
\unskip\
\newblock
\APACrefYearMonthDay{2019}{}{}.
\newblock
\APACrefbtitle {Invariant Risk Minimization.} {Invariant risk minimization.}
\newblock
\APACrefnote{arXiv:1907.02893}
\PrintBackRefs{\CurrentBib}

\bibitem [\protect \citeauthoryear {%
{B{\"u}hlmann}%
, Peters%
\BCBL {}\ \BBA {} Ernest%
}{%
{B{\"u}hlmann}%
\ \protect \BOthers {.}}{%
{\protect \APACyear {2014}}%
}]{%
Buehlmann14}
\APACinsertmetastar {%
Buehlmann14}%
\begin{APACrefauthors}%
{B{\"u}hlmann}, P.%
, Peters, J.%
\BCBL {}\ \BBA {} Ernest, J.%
\end{APACrefauthors}%
\unskip\
\newblock
\APACrefYearMonthDay{2014}{}{}.
\newblock
{\BBOQ}\APACrefatitle {{CAM}: Causal Additive Models, high-dimensional order
  search and penalized regression} {{CAM}: Causal additive models,
  high-dimensional order search and penalized regression}.{\BBCQ}
\newblock
\APACjournalVolNumPages{Ann.\ Stat.}{42}{}{2526-2556}.
\PrintBackRefs{\CurrentBib}

\bibitem [\protect \citeauthoryear {%
Chickering%
}{%
Chickering%
}{%
{\protect \APACyear {2002}}%
{\protect \APACexlab {{\protect \BCnt {1}}}}}]{%
Chickering2002}
\APACinsertmetastar {%
Chickering2002}%
\begin{APACrefauthors}%
Chickering, D\BPBI M.%
\end{APACrefauthors}%
\unskip\
\newblock
\APACrefYearMonthDay{2002{\protect \BCnt {1}}}{}{}.
\newblock
{\BBOQ}\APACrefatitle {Learning equivalence classes of {B}ayesian-network
  structures} {Learning equivalence classes of {B}ayesian-network
  structures}.{\BBCQ}
\newblock
\APACjournalVolNumPages{J. Mach. Learn. Res.}{2}{}{445--498}.
\PrintBackRefs{\CurrentBib}

\bibitem [\protect \citeauthoryear {%
Chickering%
}{%
Chickering%
}{%
{\protect \APACyear {2002}}%
{\protect \APACexlab {{\protect \BCnt {2}}}}}]{%
ChickeringGES}
\APACinsertmetastar {%
ChickeringGES}%
\begin{APACrefauthors}%
Chickering, D\BPBI M.%
\end{APACrefauthors}%
\unskip\
\newblock
\APACrefYearMonthDay{2002{\protect \BCnt {2}}}{}{}.
\newblock
{\BBOQ}\APACrefatitle {Optimal Structure Identification With Greedy Search}
  {Optimal structure identification with greedy search}.{\BBCQ}
\newblock
\APACjournalVolNumPages{J. Mach. Learn. Res.}{3}{}{507-554}.
\PrintBackRefs{\CurrentBib}

\bibitem [\protect \citeauthoryear {%
Claassen%
, Mooij%
\BCBL {}\ \BBA {} Heskes%
}{%
Claassen%
\ \protect \BOthers {.}}{%
{\protect \APACyear {2013}}%
}]{%
Claassen2013}
\APACinsertmetastar {%
Claassen2013}%
\begin{APACrefauthors}%
Claassen, T.%
, Mooij, J\BPBI M.%
\BCBL {}\ \BBA {} Heskes, T.%
\end{APACrefauthors}%
\unskip\
\newblock
\APACrefYearMonthDay{2013}{}{}.
\newblock
{\BBOQ}\APACrefatitle {Learning sparse causal models is not {NP}-hard}
  {Learning sparse causal models is not {NP}-hard}.{\BBCQ}
\newblock
\BIn{} A.~Nicholson\ \BBA {} P.~Smyth\ (\BEDS), \APACrefbtitle {Proceedings of
  {UAI} 2013} {Proceedings of {UAI} 2013}\ (\BPG~172–181).
\newblock
\APACaddressPublisher{}{AUAI Press}.
\PrintBackRefs{\CurrentBib}

\bibitem [\protect \citeauthoryear {%
Colombo%
\ \BBA {} Maathuis%
}{%
Colombo%
\ \BBA {} Maathuis%
}{%
{\protect \APACyear {2014}}%
}]{%
ColomboMaathuis2014}
\APACinsertmetastar {%
ColomboMaathuis2014}%
\begin{APACrefauthors}%
Colombo, D.%
\BCBT {}\ \BBA {} Maathuis, M\BPBI H.%
\end{APACrefauthors}%
\unskip\
\newblock
\APACrefYearMonthDay{2014}{}{}.
\newblock
{\BBOQ}\APACrefatitle {Order-Independent Constraint-Based Causal Structure
  Learning} {Order-independent constraint-based causal structure
  learning}.{\BBCQ}
\newblock
\APACjournalVolNumPages{J.\ Mach.\ Learn.\ Res.}{15}{}{3921-3962}.
\PrintBackRefs{\CurrentBib}

\bibitem [\protect \citeauthoryear {%
Colombo%
, Maathuis%
, Kalisch%
\BCBL {}\ \BBA {} Richardson%
}{%
Colombo%
\ \protect \BOthers {.}}{%
{\protect \APACyear {2012}}%
}]{%
Colombo2012}
\APACinsertmetastar {%
Colombo2012}%
\begin{APACrefauthors}%
Colombo, D.%
, Maathuis, M\BPBI H.%
, Kalisch, M.%
\BCBL {}\ \BBA {} Richardson, T\BPBI S.%
\end{APACrefauthors}%
\unskip\
\newblock
\APACrefYearMonthDay{2012}{}{}.
\newblock
{\BBOQ}\APACrefatitle {Learning high-dimensional directed acyclic graphs with
  latent and selection variables} {Learning high-dimensional directed acyclic
  graphs with latent and selection variables}.{\BBCQ}
\newblock
\APACjournalVolNumPages{Ann.\ Stat.}{40}{}{294-321}.
\PrintBackRefs{\CurrentBib}

\bibitem [\protect \citeauthoryear {%
Eigenmann%
, Nandy%
\BCBL {}\ \BBA {} Maathuis%
}{%
Eigenmann%
\ \protect \BOthers {.}}{%
{\protect \APACyear {2017}}%
}]{%
Eigenmann2017}
\APACinsertmetastar {%
Eigenmann2017}%
\begin{APACrefauthors}%
Eigenmann, M\BPBI F.%
, Nandy, P.%
\BCBL {}\ \BBA {} Maathuis, M\BPBI H.%
\end{APACrefauthors}%
\unskip\
\newblock
\APACrefYearMonthDay{2017}{}{}.
\newblock
{\BBOQ}\APACrefatitle {Structure Learning of linear {G}aussian Structural
  Equation Models with Weak Edges} {Structure learning of linear {G}aussian
  structural equation models with weak edges}.{\BBCQ}
\newblock
\BIn{} G.~Elidan, K.~Kersting\BCBL {}\ \BBA {} A\BPBI T.~Ihler\ (\BEDS),
  \APACrefbtitle {Proceedings of {UAI} 2017.} {Proceedings of {UAI} 2017.}
\newblock
\APACaddressPublisher{}{AUAI Press}.
\PrintBackRefs{\CurrentBib}

\bibitem [\protect \citeauthoryear {%
{Frot}%
, {Nandy}%
\BCBL {}\ \BBA {} {Maathuis}%
}{%
{Frot}%
\ \protect \BOthers {.}}{%
{\protect \APACyear {2019}}%
}]{%
Frot2017}
\APACinsertmetastar {%
Frot2017}%
\begin{APACrefauthors}%
{Frot}, B.%
, {Nandy}, P.%
\BCBL {}\ \BBA {} {Maathuis}, M\BPBI H.%
\end{APACrefauthors}%
\unskip\
\newblock
\APACrefYearMonthDay{2019}{}{}.
\newblock
{\BBOQ}\APACrefatitle {Robust causal structure learning with some hidden
  variables} {Robust causal structure learning with some hidden
  variables}.{\BBCQ}
\newblock
\APACjournalVolNumPages{J.\ Roy.\ Stat.\ Soc.\ B}{81}{}{459–487}.
\PrintBackRefs{\CurrentBib}

\bibitem [\protect \citeauthoryear {%
Harris%
\ \BBA {} Drton%
}{%
Harris%
\ \BBA {} Drton%
}{%
{\protect \APACyear {2013}}%
}]{%
Harris2013}
\APACinsertmetastar {%
Harris2013}%
\begin{APACrefauthors}%
Harris, N.%
\BCBT {}\ \BBA {} Drton, M.%
\end{APACrefauthors}%
\unskip\
\newblock
\APACrefYearMonthDay{2013}{}{}.
\newblock
{\BBOQ}\APACrefatitle {{PC} Algorithm for Nonparanormal Graphical Models} {{PC}
  algorithm for nonparanormal graphical models}.{\BBCQ}
\newblock
\APACjournalVolNumPages{J. Mach. Learn. Res.}{14}{}{3365-3383}.
\PrintBackRefs{\CurrentBib}

\bibitem [\protect \citeauthoryear {%
Hauser%
\ \BBA {} B{\"u}hlmann%
}{%
Hauser%
\ \BBA {} B{\"u}hlmann%
}{%
{\protect \APACyear {2012}}%
}]{%
Hauser2012}
\APACinsertmetastar {%
Hauser2012}%
\begin{APACrefauthors}%
Hauser, A.%
\BCBT {}\ \BBA {} B{\"u}hlmann, P.%
\end{APACrefauthors}%
\unskip\
\newblock
\APACrefYearMonthDay{2012}{}{}.
\newblock
{\BBOQ}\APACrefatitle {Characterization and Greedy Learning of Interventional
  {M}arkov Equivalence Classes of Directed Acyclic Graphs} {Characterization
  and greedy learning of interventional {M}arkov equivalence classes of
  directed acyclic graphs}.{\BBCQ}
\newblock
\APACjournalVolNumPages{J. Mach. Learn. Res.}{13}{}{2409--2464}.
\PrintBackRefs{\CurrentBib}

\bibitem [\protect \citeauthoryear {%
{Heinze-Deml}%
, {Maathuis}%
\BCBL {}\ \BBA {} {Meinshausen}%
}{%
{Heinze-Deml}%
\ \protect \BOthers {.}}{%
{\protect \APACyear {2018}}%
}]{%
Heinze2018}
\APACinsertmetastar {%
Heinze2018}%
\begin{APACrefauthors}%
{Heinze-Deml}, C.%
, {Maathuis}, M\BPBI H.%
\BCBL {}\ \BBA {} {Meinshausen}, N.%
\end{APACrefauthors}%
\unskip\
\newblock
\APACrefYearMonthDay{2018}{}{}.
\newblock
{\BBOQ}\APACrefatitle {{Causal Structure Learning}} {{Causal Structure
  Learning}}.{\BBCQ}
\newblock
\APACjournalVolNumPages{Annu.\ Rev.\ Stat.\ Appl.}{5}{}{371-391}.
\PrintBackRefs{\CurrentBib}

\bibitem [\protect \citeauthoryear {%
Hill%
, Heiser%
, Cokelaer%
\BCBL {}\ \BBA {} et al.%
}{%
Hill%
\ \protect \BOthers {.}}{%
{\protect \APACyear {2016}}%
}]{%
Hill2016}
\APACinsertmetastar {%
Hill2016}%
\begin{APACrefauthors}%
Hill, S\BPBI M.%
, Heiser, L.%
, Cokelaer, T.%
\BCBL {}\ \BBA {} et al.%
\end{APACrefauthors}%
\unskip\
\newblock
\APACrefYearMonthDay{2016}{}{}.
\newblock
{\BBOQ}\APACrefatitle {Inferring causal molecular networks: Empirical
  assessment through a community-based effort} {Inferring causal molecular
  networks: Empirical assessment through a community-based effort}.{\BBCQ}
\newblock
\APACjournalVolNumPages{Nat.\ Methods}{13}{}{310--318}.
\PrintBackRefs{\CurrentBib}

\bibitem [\protect \citeauthoryear {%
Hill%
, Oates%
, Blythe%
\BCBL {}\ \BBA {} Mukherjee%
}{%
Hill%
\ \protect \BOthers {.}}{%
{\protect \APACyear {2019}}%
}]{%
Hill2019}
\APACinsertmetastar {%
Hill2019}%
\begin{APACrefauthors}%
Hill, S\BPBI M.%
, Oates, C\BPBI J.%
, Blythe, D\BPBI A.%
\BCBL {}\ \BBA {} Mukherjee, S.%
\end{APACrefauthors}%
\unskip\
\newblock
\APACrefYearMonthDay{2019}{}{}.
\newblock
{\BBOQ}\APACrefatitle {Causal Learning via Manifold Regularization} {Causal
  learning via manifold regularization}.{\BBCQ}
\newblock
\APACjournalVolNumPages{J.\ Mach.\ Learn.\ Res.}{20}{}{1-32}.
\PrintBackRefs{\CurrentBib}

\bibitem [\protect \citeauthoryear {%
Kalisch%
, M\"achler%
, Colombo%
, Maathuis%
\BCBL {}\ \BBA {} B\"uhlmann%
}{%
Kalisch%
\ \protect \BOthers {.}}{%
{\protect \APACyear {2012}}%
}]{%
Kalisch2012}
\APACinsertmetastar {%
Kalisch2012}%
\begin{APACrefauthors}%
Kalisch, M.%
, M\"achler, M.%
, Colombo, D.%
, Maathuis, M\BPBI H.%
\BCBL {}\ \BBA {} B\"uhlmann, P.%
\end{APACrefauthors}%
\unskip\
\newblock
\APACrefYearMonthDay{2012}{}{}.
\newblock
{\BBOQ}\APACrefatitle {Causal Inference Using Graphical Models with the {R}
  Package {pcalg}} {Causal inference using graphical models with the {R}
  package {pcalg}}.{\BBCQ}
\newblock
\APACjournalVolNumPages{J.\ Stat.\ Softw.}{47}{}{11: 1--26}.
\PrintBackRefs{\CurrentBib}

\bibitem [\protect \citeauthoryear {%
Maathuis%
, Kalisch%
\BCBL {}\ \BBA {} Bühlmann%
}{%
Maathuis%
\ \protect \BOthers {.}}{%
{\protect \APACyear {2009}}%
}]{%
Maathuis2009}
\APACinsertmetastar {%
Maathuis2009}%
\begin{APACrefauthors}%
Maathuis, M\BPBI H.%
, Kalisch, M.%
\BCBL {}\ \BBA {} Bühlmann, P.%
\end{APACrefauthors}%
\unskip\
\newblock
\APACrefYearMonthDay{2009}{}{}.
\newblock
{\BBOQ}\APACrefatitle {Estimating high-dimensional intervention effects from
  observational data} {Estimating high-dimensional intervention effects from
  observational data}.{\BBCQ}
\newblock
\APACjournalVolNumPages{Ann.\ Stat.}{37}{}{3133--3164}.
\PrintBackRefs{\CurrentBib}

\bibitem [\protect \citeauthoryear {%
Mooij%
\ \BBA {} Heskes%
}{%
Mooij%
\ \BBA {} Heskes%
}{%
{\protect \APACyear {2013}}%
}]{%
Mooij2013}
\APACinsertmetastar {%
Mooij2013}%
\begin{APACrefauthors}%
Mooij, J\BPBI M.%
\BCBT {}\ \BBA {} Heskes, T.%
\end{APACrefauthors}%
\unskip\
\newblock
\APACrefYearMonthDay{2013}{}{}.
\newblock
{\BBOQ}\APACrefatitle {Cyclic Causal Discovery from Continuous Equilibrium
  Data} {Cyclic causal discovery from continuous equilibrium data}.{\BBCQ}
\newblock
\BIn{} A.~Nicholson\ \BBA {} P.~Smyth\ (\BEDS), \APACrefbtitle {Proceedings of
  {UAI} 2013} {Proceedings of {UAI} 2013}\ (\BPGS\ 431--439).
\newblock
\APACaddressPublisher{}{AUAI Press}.
\PrintBackRefs{\CurrentBib}

\bibitem [\protect \citeauthoryear {%
{Nandy}%
, {Hauser}%
\BCBL {}\ \BBA {} {Maathuis}%
}{%
{Nandy}%
\ \protect \BOthers {.}}{%
{\protect \APACyear {2018}}%
}]{%
Nandy2015}
\APACinsertmetastar {%
Nandy2015}%
\begin{APACrefauthors}%
{Nandy}, P.%
, {Hauser}, A.%
\BCBL {}\ \BBA {} {Maathuis}, M\BPBI H.%
\end{APACrefauthors}%
\unskip\
\newblock
\APACrefYearMonthDay{2018}{}{}.
\newblock
{\BBOQ}\APACrefatitle {High-dimensional consistency in score-based and hybrid
  structure learning} {High-dimensional consistency in score-based and hybrid
  structure learning}.{\BBCQ}
\newblock
\APACjournalVolNumPages{Ann.\ Stat.}{46}{}{3151-3183}.
\PrintBackRefs{\CurrentBib}

\bibitem [\protect \citeauthoryear {%
Peters%
\ \BBA {} B{\"u}hlmann%
}{%
Peters%
\ \BBA {} B{\"u}hlmann%
}{%
{\protect \APACyear {2015}}%
}]{%
Peters2015}
\APACinsertmetastar {%
Peters2015}%
\begin{APACrefauthors}%
Peters, J.%
\BCBT {}\ \BBA {} B{\"u}hlmann, P.%
\end{APACrefauthors}%
\unskip\
\newblock
\APACrefYearMonthDay{2015}{}{}.
\newblock
{\BBOQ}\APACrefatitle {Structural Intervention Distance for Evaluating Causal
  Graphs} {Structural intervention distance for evaluating causal
  graphs}.{\BBCQ}
\newblock
\APACjournalVolNumPages{Neural Comput.}{27}{}{771-799}.
\PrintBackRefs{\CurrentBib}

\bibitem [\protect \citeauthoryear {%
Peters%
, B{\"u}hlmann%
\BCBL {}\ \BBA {} Meinshausen%
}{%
Peters%
\ \protect \BOthers {.}}{%
{\protect \APACyear {2016}}%
}]{%
Peters2015b}
\APACinsertmetastar {%
Peters2015b}%
\begin{APACrefauthors}%
Peters, J.%
, B{\"u}hlmann, P.%
\BCBL {}\ \BBA {} Meinshausen, N.%
\end{APACrefauthors}%
\unskip\
\newblock
\APACrefYearMonthDay{2016}{}{}.
\newblock
{\BBOQ}\APACrefatitle {Causal inference using invariant prediction:
  identification and confidence intervals} {Causal inference using invariant
  prediction: identification and confidence intervals}.{\BBCQ}
\newblock
\APACjournalVolNumPages{J.\ Roy.\ Stat.\ Soc.\ B}{78}{}{947--1012}.
\PrintBackRefs{\CurrentBib}

\bibitem [\protect \citeauthoryear {%
Sachs%
, Perez%
, Pe'er%
, Lauffenburger%
\BCBL {}\ \BBA {} Nolan%
}{%
Sachs%
\ \protect \BOthers {.}}{%
{\protect \APACyear {2005}}%
}]{%
Sachs2005}
\APACinsertmetastar {%
Sachs2005}%
\begin{APACrefauthors}%
Sachs, K.%
, Perez, O.%
, Pe'er, D.%
, Lauffenburger, D\BPBI A.%
\BCBL {}\ \BBA {} Nolan, G\BPBI P.%
\end{APACrefauthors}%
\unskip\
\newblock
\APACrefYearMonthDay{2005}{}{}.
\newblock
{\BBOQ}\APACrefatitle {{Causal protein-signaling networks derived from
  multiparameter single-cell data}} {{Causal protein-signaling networks derived
  from multiparameter single-cell data}}.{\BBCQ}
\newblock
\APACjournalVolNumPages{Science}{308}{}{523--529}.
\PrintBackRefs{\CurrentBib}

\bibitem [\protect \citeauthoryear {%
Shimizu%
, Hoyer%
, Hyv{\"a}rinen%
\BCBL {}\ \BBA {} Kerminen%
}{%
Shimizu%
\ \protect \BOthers {.}}{%
{\protect \APACyear {2006}}%
}]{%
Shimizu2006}
\APACinsertmetastar {%
Shimizu2006}%
\begin{APACrefauthors}%
Shimizu, S.%
, Hoyer, P.%
, Hyv{\"a}rinen, A.%
\BCBL {}\ \BBA {} Kerminen, A.%
\end{APACrefauthors}%
\unskip\
\newblock
\APACrefYearMonthDay{2006}{}{}.
\newblock
{\BBOQ}\APACrefatitle {A linear non-{G}aussian acyclic model for causal
  discovery} {A linear non-{G}aussian acyclic model for causal
  discovery}.{\BBCQ}
\newblock
\APACjournalVolNumPages{J. Mach. Learn. Res.}{7}{}{2003--2030}.
\PrintBackRefs{\CurrentBib}

\bibitem [\protect \citeauthoryear {%
Spirtes%
, Glymour%
\BCBL {}\ \BBA {} Scheines%
}{%
Spirtes%
\ \protect \BOthers {.}}{%
{\protect \APACyear {2000}}%
}]{%
Spirtes2000}
\APACinsertmetastar {%
Spirtes2000}%
\begin{APACrefauthors}%
Spirtes, P.%
, Glymour, C.%
\BCBL {}\ \BBA {} Scheines, R.%
\end{APACrefauthors}%
\unskip\
\newblock
\APACrefYear{2000}.
\newblock
\APACrefbtitle {Causation, Prediction, and Search} {Causation, prediction, and
  search}\ (\PrintOrdinal{Second}\ \BEd).
\newblock
\APACaddressPublisher{}{MIT Press, Cambridge}.
\newblock
\APACrefnote{With additional material by D.\ Heckerman, C.\ Meek, G.F.\ Cooper
  and T.\ Richardson}
\PrintBackRefs{\CurrentBib}

\bibitem [\protect \citeauthoryear {%
Tsamardinos%
, Brown%
\BCBL {}\ \BBA {} Aliferis%
}{%
Tsamardinos%
\ \protect \BOthers {.}}{%
{\protect \APACyear {2006}}%
}]{%
Tsamardinos2006}
\APACinsertmetastar {%
Tsamardinos2006}%
\begin{APACrefauthors}%
Tsamardinos, I.%
, Brown, L\BPBI E.%
\BCBL {}\ \BBA {} Aliferis, C\BPBI F.%
\end{APACrefauthors}%
\unskip\
\newblock
\APACrefYearMonthDay{2006}{}{}.
\newblock
{\BBOQ}\APACrefatitle {The Max-min Hill-climbing {B}ayesian Network Structure
  Learning Algorithm} {The max-min hill-climbing {B}ayesian network structure
  learning algorithm}.{\BBCQ}
\newblock
\APACjournalVolNumPages{Mach. Learn.}{65}{}{31--78}.
\PrintBackRefs{\CurrentBib}

\end{thebibliography}

\end{document}